\documentclass[]{piparticle-final}
\usepackage{graphicx}
\usepackage{amsmath}

\usepackage{cite} % To compress citation ranges
\usepackage{epstopdf}

\begin{document}

\volume{5}               % To be inserted by Editor
\articlenumber{050010}   % To be inserted by Editor
\journalyear{2013}       % To be inserted by Editor
\editor{R. Dickman}   % To be inserted by Editor
\reviewers{F. Reis, Instituto de F\'isica, Univ. Fed. Fluminense, Brazil.}  % To be inserted by Editor
\received{25 June 2013}     % To be inserted by Editor
\accepted{10 December 2013}   % To be inserted by Editor
\runningauthor{H. S. Wio \itshape{et al.}}  % To be inserted by Editor
\doi{050010}         % To be inserted by Editor

\title{Invited review: KPZ. Recent developments via a variational formulation}

% Institution references with \cite are inserted after \maketitle in theaffiliation enviroment
\author{Horacio S. Wio,\cite{inst1}\thanks{E-mail: wio@ifca.unican.es} \hspace{0.4em}
Roberto R. Deza,\cite{inst2}\thanks{E-mail: deza@mdp.edu.ar} \hspace{0.4em}
Carlos Escudero,\cite{inst3}\thanks{E-mail: cel@icmat.es} \hspace{0.4em}
Jorge A. Revelli\cite{inst4}\thanks{E-mail: revelli@famaf.unc.edu.ar}
}

\pipabstract{
Recently, a variational approach has been introduced for the paradigmatic Kardar--Parisi--Zhang
(KPZ) equation. Here we review that approach, together with the functional Taylor expansion that
the KPZ nonequilibrium potential (NEP) admits. Such expansion becomes naturally truncated at
third order, giving rise to a nonlinear stochastic partial differential equation to be regarded
as a gradient-flow counterpart to the KPZ equation. A dynamic renormalization group analysis
at one-loop order of this new mesoscopic model yields the KPZ scaling relation $\alpha+z=2$, as
a consequence of the exact cancelation of the different contributions to vertex renormalization.
This result is quite remarkable, considering the lower degree of symmetry of this equation, which
is in particular not Galilean invariant. In addition, this scheme is exploited to
inquire about the dynamical behavior of the KPZ equation through a path-integral approach. Each of
these aspects offers novel points of view and sheds light on particular aspects of the dynamics
of the KPZ equation.
}

\maketitle

\blfootnote{
\begin{theaffiliation}{99}
   \institution{inst1} IFCA (UC-CSIC), Avda. de los Castros s/n, E-39005 Santander, Spain.
   \institution{inst2} IFIMAR (UNMdP-CONICET), Funes 3350, 7600 Mar del Plata, Argentina.
   \institution{inst3} Depto. Matem\'aticas \& ICMAT (CSIC-UAM-UC3M-UCM), Cantoblanco, E-28049 Madrid, Spain.
   \institution{inst4} FaMAF-IFEG (CONICET-UNC), 5000 C\'ordoba, Argentina.

\end{theaffiliation}
}

\section{Introduction}
Although readers whose careers span mostly on the 21th century might not care about this,
back in the sixties (when transistors and lasers had already been invented)
\emph{equilibrium} critical phenomena were still a puzzle. In fact, although a sense
of ``universality'' had been gained in 1950 through a field theory based on the
innovative concept of \emph{order parameter} \cite{landau,kadan60}, its predicted critical
exponents were almost as a rule wrong. It was not until the seventies that a far more
sophisticated field-theory approach \cite{wilson} brought order home: equilibrium
universality classes are determined solely by the dimensionalities of the order
parameter and the ambient space. Since then, one of Statistical Physics' ``holy grials''
has been to conquer a similar achievement for \emph{non-equilibrium} critical phenomena
\cite{marro}.
In such a (still unaccomplished) enterprise, a valuable field-theoretical tool has
been in the last quarter of century the Kardar--Parisi--Zhang (KPZ) equation
\cite{kpz,HHZ,BarSta}.

The KPZ equation~\cite{kpz,HHZ,BarSta} has become a paradigm for the description of a vast
class of nonequilibrium phenomena by means of stochastic fields. The field $h(x,t)$, whose
evolution is governed by this stochastic nonlinear partial differential equation, describes
the height of a fluctuating interface in the context of surface-growth processes in which
it was originally formulated. From a theoretical point of view, the KPZ equation has many
interesting properties, for instance, its close relationship with the Burgers equation~\cite{burgers}
or with a diffusion equation with multiplicative noise, whose field $\phi(x,t)$ can be
interpreted as the restricted partition function of the directed polymer problem~\cite{kardar}.
Many of the efforts put in investigating the behavior of its solutions were focused on
obtaining the scaling laws and critical exponents in one or more spatial dimensions~
\cite{BecCur,MoserWolf-Discr3d,Scalerandi-etal,NewmanBray-Discr,Appert,Marinari-etal,OAF}.
However, other questions of great interest are the development of suitable
algorithms for its numerical integration~\cite{GiadaGiacomettiRossi,Aarao1},
the construction of particular solutions~\cite{SaSp,SaSp2,AmCoQu,CaLeD}, the crossover
behavior between different regimes \cite{GGG,FoTo,Albano,Aarao2}, as well as
related ageing and pinning phenomena \cite{bustin1,bustin2,agePRE}.

Among all the classical theoretical developments concerning this equation~\cite{BarSta,HHZ}, two
have recently drawn our attention. One was the scaling relation $\alpha+z=2$, which is
expected to be exact for the KPZ equation in any dimension. The exactness of this relation has
been traditionally attributed to the Galilean invariance of the KPZ equation. Nevertheless, the
assumed central role of this symmetry has been challenged in this as well as in other
nonequilibrium models from both a theoretical~\cite{McComb,BeHo-1,BeHo-2,NiCuCa} and a numerical~
\cite{Nos-1,Nos-2,Nos-3} point of view. The second one is the generally accepted lack of existence
of a suitable functional allowing to formulate the KPZ equation as a gradient flow.
In fact, a variational approach to the closely related Sun-Guo-Grant~\cite{sun} and
Villain-Lai-Das Sarma~\cite{villain,lai} equations was developed in~\cite{escudero,elka}
by means of a geometric construction.
In~\cite{wio-01}, a Lyapunov functional (with an explicit density) was found for the
deterministic KPZ equation. Also, a \emph{nonequilibrium potential} (NEP), a functional
that allows the formal writing of the KPZ equation as a (stochastically forced) exact
gradient flow, was introduced.

In this work we shortly review the consistency constraints imposed by the nonequilibrium-potential
structure on \emph{discrete} representations of the KPZ equation and show that they
lead to explicit breakdown of Galilean invariance, despite the fact that the obtained numerical
results are still those of the KPZ universality class. A Taylor expansion of the
previously introduced NEP has (in terms of \emph{fluctuations}) an explicit density and
a thought-provoking structure \cite{Nos-PLA}, and leads to an equation of motion (for
\emph{fluctuations}, in the continuum) with exact gradient-flow structure, but different
from the KPZ one. This equation has a lower degree of symmetry: it is neither Galilean
invariant nor even translation invariant. Its scaling properties are studied by means
of a dynamic renormalization group (DRG) analysis, and its critical exponents fulfill at
one-loop order the same scaling relation $\alpha + z = 2$ as those of the KPZ equation,
despite the aforementioned lack of Galilean invariance. The concern with stability leads
us to suggest the introduction of an equation related to the Kuramoto–-Sivashinsky one,
also with exact gradient-flow structure. We close this article exposing some novel
developments based on a path-integral-like approach.

\section{Brief review of the nonequilibrium potential scheme}

Loosely speaking, the notion of NEP is an extension to
nonequilibrium situations of that of equilibrium thermodynamic
potential. In order to introduce it, we consider a general system
of nonlinear stochastic equations (admitting the possibility of
\emph{multiplicative noises})
\begin{equation}\label{eq:14}
\dot{q}^\nu=K^\nu(q)+g^\nu_i(q)\,\xi_i(t),\qquad\nu=1,\ldots,n;
\end{equation}
where repeated indices are summed over. Equation (\ref{eq:14}) is
stated in the sense of It\^o. The \(\{\xi_i(t)\}\),
\(i=1,\ldots,m\leq n\) are mutually independent sources of Gaussian
white noise with typical strength \(\gamma\).

The Fokker--Planck equation corresponding to Eq. (\ref{eq:14})
takes the form
\begin{equation}\label{eq:15}
\frac{\partial P}{\partial t}=-\frac{\partial}{\partial q^\nu}
\,K^\nu(q)\,P+\frac{\gamma}{2}\frac{\partial^2}{\partial q^\nu
\,\partial q^\mu}\,Q^{\nu\mu}(q)\,P
\end{equation}
where \(P(q,t;\gamma)\) is the probability density of observing
\(q=(q_1,\ldots,q_n)\) at time \(t\) for noise intensity \(\gamma\),
and \(Q^{\nu\mu}(q)=g^\nu_i(q)\,g^\mu_i(q)\) is the matrix
of transport coefficients of the system, which is symmetric and
non-negative. In the long time limit (\(t\to\infty\)), the solution
of Eq. (\ref{eq:15}) tends to the stationary distribution
\(P_\mathrm{st}(q)\). According to \cite{GR,hsw,wdl}, the NEP
\(\Phi(q)\) associated to Eq. (\ref{eq:15}) is defined by
\begin{equation}\label{eq:16}
\Phi(q)=-\lim_{\gamma\to0}\gamma\,\ln P_\mathrm{st}(q,\gamma).
\end{equation}
In other words, \[P_\mathrm{st}(q)\,d^nq=Z(q)\exp
\left[-\frac{\Phi(q)}{\gamma}+\mathcal{O}(\gamma)\right]\,d\Omega_q,\]
where \(\Phi(q)\) is the NEP of the system and the prefactor \(Z(q)\)
is defined as the limit \[\ln Z(q)=\lim_{\gamma\to0}
\left[\ln P_\mathrm{st}(q,\gamma)+\frac{1}{\gamma}\,\Phi(q)\right].\]
Here \(d\Omega_q=d^nq/\sqrt{G(q)}\) is the invariant volume element
in the \(q\)-space and \(G(q)\) is the determinant of the
contravariant metric tensor (for the Euclidean metric it is
\(G=1\)). It was shown \cite{GR} that \(\Phi(q)\) is the solution
of a Hamilton--Jacobi-like equation (HJE)
\[K^\nu(q)\frac{\partial\Phi}{\partial q^\nu}+
\frac{1}{2}Q^{\nu\mu}(q)\frac{\partial\Phi}{\partial
q^\nu}\frac{\partial\Phi}{\partial q^\mu}=0,\] and \(Z(q)\) is
the solution of a linear first-order partial differential equation
depending on \(\Phi(q)\) (not shown here).

Equation (\ref{eq:16}) and the normalization condition ensure that
\(\Phi\) is bounded from below. Furthermore, it follows that
\[\frac{d\Phi(q)}{dt}=K^\nu(q)
\frac{\partial\Phi(q)}{\partial q^\nu}=
-\frac{1}{2}\,Q^{\nu\mu}(q)\,\frac{\partial\Phi}
{\partial q^\nu}\frac{\partial\Phi}{\partial q^\mu}\leq 0,\]
i.e., \(\Phi\) is a Lyapunov functional for the dynamics of the system
when fluctuations are neglected. Under the deterministic dynamics,
\(\dot{q}^\nu=K^\nu(q)\), \(\Phi\) decreases monotonically and
takes a minimum value on attractors. In particular, \(\Phi\) must
be constant on all extended attractors (such as limit cycles or
strange attractors) \cite{GR}.

An alternative way to look into this problem is due to Ao \cite{pao}.
The interesting feature of this approach is that it resorts neither
to $P_{st}(q)$ nor to the small-noise limit, thus being applicable
in principle to more general situations.

\section{Variational approach for KPZ}

The Kardar--Parisi--Zhang (KPZ) equation reads
\begin{equation}\label{eq-000}
\frac{\partial h(x,t)}{\partial t}=\nu\nabla^2h(x,t)+
\frac{\lambda}{2}\left[\nabla h(x,t)\right]^2+\xi(x,t),
\end{equation}
where $\xi(x,t)$ is a Gaussian white noise, of zero mean ($\langle\xi(x,t)\rangle=0$) and
correlation $\langle\xi(x,t)\xi(x',t')\rangle= 2 \gamma \delta (x-x')\delta(t-t')$. As it is well
known, this nonlinear differential equation describes the fluctuations of a growing
interface with a surface tension given by $\nu$; $\lambda$ is proportional to the average
growth velocity and arises because the surface slope is paralleled transported in such a
growth process.

\subsection*{Lyapunov functional}
The deterministic KPZ equation---obtained by setting $\gamma =0$---is exactly solvable by means
of the Hopf--Cole transformation ($\phi(x,t) = e^{\frac{\lambda}{2 \nu} h(x,t} $), which
maps the nonlinear KPZ equation onto the (deterministic) linear diffusion equation~\cite{kpz}
\begin{equation}\label{eq-0001}
\frac{\partial \phi(x,t)}{\partial t}= \nu \nabla^2\phi(x,t).
\end{equation}
Also, the multiplicative reaction-diffusion (RD) equation
\begin{equation}\label{eq-0002}
\frac{\partial \phi(x,t)}{\partial t}= \nu\nabla^2\phi(x,t) + \phi(x,t) \xi(x,t),
\end{equation}
which is associated to the directed polymer problem \cite{HHZ,BarSta,kardar} results,
using the inverse transformation ($h(x,t) = \frac{2 \nu}{\lambda} \ln \phi(x,t)$)
to be mapped into the \emph{complete} KPZ equation (\ref{eq-000}).

The deterministic part of Eq. (\ref{eq-0002}) (i.e., Eq. (\ref{eq-0001})),
can be written as
\begin{equation}\label{eq-0001p}
\frac{\partial \phi(x,t)}{\partial t}= - \frac{\delta \mathcal{F}[\phi(x,t)]}{\delta \phi(x,t)},
\end{equation}
where $\mathcal{F}[\phi(x,t)]$ is the Lyapunov functional
of the deterministic RD problem given by
\[ \mathcal{F}[\phi(x,t)] = \frac{\nu}{2} \,\int
\left[\nabla \phi(x,t)\right]^2\,\mathrm{d} x.\]
Applying to this functional the
above indicated inverse transformation we get~\cite{wio-01}
\begin{equation}\label{eq-001}
\mathcal{F}[h]= \frac{\lambda ^2}{8\nu}\,\int e^{\frac{\lambda}{\nu} h(x,t)}
\left[\nabla h(x,t)\right]^2\,\mathrm{d} x,
\end{equation}
that allows the KPZ equation to be written as
\begin{equation}\label{eq-002}
\frac{\partial}{\partial t}h(x,t)=
-\Gamma[h]\frac{\delta\mathcal{F}[h]}{\delta h(x,t)}+\xi(x,t).
\end{equation}
One can check the Lyapunov property $\dot{\mathcal{F}}[h]
\leq0$, with the motility $\Gamma[h]$ given by
$$\Gamma[h]=\left(\frac{2\nu}{\lambda}\right)^2e^{-\frac{\lambda}{\nu}h(x,t)},$$
and that its minimum is achieved by constant functions. Hence we have a
Lyapunov functional for the deterministic KPZ equation that displays simple dynamics:
the asymptotic stability of constant solutions
indicates an approach to constant profiles at long times, for arbitrary initial conditions.
Despite this simplicity in the deterministic case, the stochastic situation is far from
trivial and gives rise to self-affine fractal profiles. In particular, the existence of
this Lyapunov functional provides no a priori intuition on the stochastic dynamics.

\subsection*{The nonequilibrium potential}

An alternative functional was also proposed in~\cite{wio-01}. By starting from
the functional Fokker-Planck equation, we look for the
stationary solution (in fact steady-state solution), and after some
integration by parts, it is possible to arrive to another form of
Lyapunov functional
\begin{equation}\label{nep-01}
\Phi[h]=\frac{\nu}{2}\int\,\mathrm{d} x\left(\nabla h\right)^2-\frac{\lambda}{2}
\int\,\mathrm{d} x\int_{h_\mathrm{ref}}^{h(x,t)}\mathrm{d}\psi\,\left(\nabla\psi\right)^2.
\end{equation}
It is somehow inspired in the analytical form of ``model A'', according to the
classification of critical phenomena in~\cite{hoho}. Here, the interpretation
of the integral in the 2nd term on the rhs is
$\int\,\mathrm{d} x\int_{h_\mathrm{ref}}^{h(x,t)}\mathrm{d}\psi = \sum_j  \triangle x \int_{h_\mathrm{ref,j}}^{h_j} d\psi _j.$
According to this definition, the KPZ equation can be formally written
as a stochastically forced gradient flow
\begin{equation}\label{kpz-var}
\frac{\partial}{\partial t}h(x,t)=-\frac{\delta\Phi[h]}{\delta h(x,t)}+\xi(x,t).
\end{equation}
The functional so defined fulfills the Lyapunov condition
$\dot{\Phi}[h]=-\left(\frac{\delta\Phi[h]}{\delta h(x,t)}\right)^2 \leq0$
as well, and could be identified as the \emph{nonequilibrium potential} (NEP) for the
KPZ case~\cite{WiHaBo,WiDe}.

We will not pursue here the development of rigorous result concerning the functional
(\ref{nep-01}). Our present interest falls in the calculation of quantities of
physical interest rather than in building a completely rigorous mathematical
theory. It is worth remarking that, as indicated in~\cite{wio-01} and above,
such a form has a discrete definition. It is also
interesting to point out that analogous functionals involving functional integrals
which are not carried out explicitly were obtained for the problem of interface
fluctuations in random media~\cite{grima,kole,bruaep,keletu}.

\subsection*{NEP expansion}

We now proceed to formally Taylor expand the NEP defined in Eq.~(\ref{nep-01})
around a given reference (or initial) state, denoted by $h_0$
\begin{eqnarray} \label{nep-10}
\Phi[h]&=&\frac{\nu}{2}\int\,\mathrm{d} x\left(\nabla h\right)^2 \nonumber\\
& & \,\,\,\,\,\,\,\, -\frac{\lambda}{2}
\int\,\mathrm{d} x\int_{h_\mathrm{0}}^{h(x,t)}\mathrm{d}\psi\,\left(\nabla\psi\right)^2\nonumber\\
&\approx&\Phi[h_0]+\delta\Phi[h_0]+\frac{1}{2}\,\delta^2\Phi[h_0]\nonumber\\
& & \,\,\,\,\,\,\,\,\,\,\,\,\,\,+ \frac{1}{6}\,\delta^3\Phi[h_0]+\cdots .
\end{eqnarray}
The successive terms in the expansion of $\Phi [h]$ are
\begin{eqnarray} \label{nep-11}
\delta \Phi [h_0] & = & - \int \,\mathrm{d} x \left[ \nu \, \nabla^2 h_0 +
\frac{\lambda}{2} \, \left(\nabla h_0 \right)^2 \right] \delta h,
\nonumber \\
\delta^2 \Phi [h_0] & = & - \int \,\mathrm{d} x \, \delta h \left( \nu
\nabla^2 + \, \lambda \nabla h_0 \cdot \nabla \right) \delta h,
\nonumber \\
\delta^3 \Phi [h_0] & = & - \lambda \, \int \,\mathrm{d} x \, \delta h
\left(\nabla \delta h \right)^2.
\end{eqnarray}
Clearly, for higher order ($n \geq 4$) terms we have
\begin{eqnarray} \label{nep-12}
\delta^n \Phi [h_0] & \equiv & 0,
\end{eqnarray}
indicating that this formal expansion has a natural cut-off after the third order.

It is worth indicating that in
this computation---as in all the other computations within this work---boundary terms
vanish provided one of the following types of boundary conditions is assumed:
homogeneous Dirichlet boundary conditions, homogeneous Neumann boundary conditions,
periodic boundary conditions or an infinite space with the derivatives of $\delta h$
vanishing as they approach an infinite distance from the origin.

The reference state $h_0$ is arbitrary (i.e., any initial condition), but it is
particularly useful to take it as one that makes $\delta\Phi[h_0]=0$, that is:
a solution to the stationary counterpart of the deterministic KPZ equation. The
complete set of solutions is $h_0=c$, where $c$ is an arbitrary constant (arbitrary
up to the application of the boundary conditions, whenever this consideration applies),
what physically corresponds to a flat interface. Hence we have ($\delta h = h -h_0$)
\begin{eqnarray} \label{nep-13}
\Phi [h] & = & \Phi [h_0] + \frac{1}{2} \, \delta^2 \Phi
[h_0, \delta h] + \frac{1}{6} \, \delta^3 \Phi [h_0, \delta h].\nonumber\\
\end{eqnarray}

\subsection*{The equation for fluctuations}

From here we can define an effective NEP, which drives the dynamics of the fluctuations
$\delta h$ and has an explicit density. Clearly, it corresponds to the last two terms in
Eq.~(\ref{nep-13}). To simplify the notation we adopt $u(x,t):=\delta h(x,t)$, and so the
NEP reads
\begin{eqnarray} \label{nep-17}
\mathcal{I}[u]&=&\int \,\mathrm{d} x \,\left[ \frac{\nu}{2} - \frac{\lambda}{6} \, u(x,t)
\right] \left(\nabla u \right)^2 \\
 &=& - \int \mathrm{d} x \, u(x,t) \left[ \nu \nabla ^2 u + \frac{\lambda}{6}
\left(\nabla u \right)^2 \right]. \nonumber
\end{eqnarray}

The deterministic equation for $u$ results
\begin{eqnarray} \label{nep-18}
\frac{\partial u}{\partial t} & = & - \frac{\delta
\mathcal{I}[u]}{\delta u},\nonumber \\
\frac{\partial u}{\partial t} & = & \left( \nu
-\frac{\lambda u}{3} \right) \nabla^2 u - \frac{\lambda}{6} (\nabla u)^2.
\end{eqnarray}
Clearly, patterns like $u_0=$ constant are stationary solutions of Eq.~(\ref{nep-18}): for all
of them $\mathcal{I} [u]=0$, indicating that all such states have the same ``energy". Finally,
let us remark that although the formal Taylor expansion becomes naturally truncated at third
order, the deterministic KPZ equation is not recovered. We call the stochastic version of
this new equation ``KPZW".
%This is not surprising because nothing
%guarantees the analyticity of such a construction.

There is a remarkable difference between both equations (KPZ and
Eq. (\ref{nep-18})). It arises due to the fact that in the first
case we have a \textbf{fixed} equation for $h$ and for \textbf{any} initial
condition, while in the second case we have a \textbf{fixed initial condition}
($u=0$) with a \textbf{variable equation} whose coefficients depend on $h_o$!
(it is an equation for the departure from the given initial condition).
The question of the relevance of this aspect to ageing problems (as discussed for 
instance in \cite{agePRE}) arises naturally. This point, worth to be analyzed, 
will be the subject of further work.

\subsection*{Non-local kernel}
In previous works \cite{Nos-3,wio-01} it was indicated that the
following functional, including a nonlocal contribution,
\begin{eqnarray}
\mathcal{F}[h]&=&\int_\Omega \left\{
\left(\frac{\lambda^2}{8\nu}\right)\left(\nabla h \right)^2+
e^{-\frac{\lambda}{2\nu}h(\mathbf{x},t)}\right. \nonumber\\
& & \times \left. \int_\Omega d\mathbf{x}'G(\mathbf{x},\mathbf{x}')
e^{\frac{\lambda}{2\nu}h(\mathbf{x}',t)} \right\}
e^{\frac{\lambda}{\nu}h(\mathbf{x},t)} d\mathbf{x},\nonumber\\
& &
\label{eq:4.01}
\end{eqnarray}
leads, after functional derivation, to a generalized KPZ equation
\begin{eqnarray}
\partial_th(\mathbf{x},t)&=&\nu\nabla^2
h(\mathbf{x},t)+\frac{\lambda}{2}[\nabla h(\mathbf{x},t)]^2 \nonumber\\
&&-  e^{-\frac{\lambda}{2\nu}h(\mathbf{x},t)}
\int_\Omega d\mathbf{x}'G(\mathbf{x},\mathbf{x}')
e^{\frac{\lambda}{2\nu}h(\mathbf{x}',t)}\nonumber\\
& & \,\,\,\,\,\,\,\,\,\,\,\, +\xi(\mathbf{x},t).
\label{eq:4.02}
\end{eqnarray}
It was also shown that if the nonlocal kernel has translational invariance
(\(G(\mathbf{x},\mathbf{x}')=G(\mathbf{x}-\mathbf{x}')\)), and
also, if it is of (very) ``short'' range, it can be expanded as
\begin{equation}\label{eq:4.03}
G(\mathbf{x}-\mathbf{x}')=\sum_{n=0}^\infty A_{2n}
\delta^{(2n)}(\mathbf{x}-\mathbf{x}'),
\end{equation}
with \(\delta^{(n)}(\mathbf{x}-\mathbf{x}')=\nabla_{\mathbf{x}'}^{n}
\delta(\mathbf{x}-\mathbf{x}')\), and where symmetry properties were taken into account.
Exploiting this form of the kernel, and considering different approximation
orders, it is possible to recover contributions having the same form as the ones
arising in several previous works, where scaling properties, symmetry
arguments, etc., have been used to discuss the possible contributions
to a general form of the kinetic equation \cite{hent94,lirh00,locg05}.
Such different contributions are tightly related to several of other previously
studied equations, like the Sun--Guo--Grant equation \cite{sun}, as well
as others \cite{hent94,caea07}.

We will not pursue this aspect here, but we will briefly refer again to it in
a forthcoming section.

\section{\label{sec:3} Discretization issues, symmetry violation and
all that}

In this section we will review aspects related to two main symmetries
associated with the 1D KPZ
equation: Galilean invariance and the fluctuation--dissipation
relation. On the one hand, Galilean invariance has been traditionally
linked to the exactness of the relation \(\alpha+z=2\) among the
critical exponents, in any spatial dimensionality (the roughness
exponent \(\alpha\), characterizing the surface morphology in the
stationary regime, and the dynamic exponent \(z\), indicating the
correlation length scaling as \(\xi(t)\sim t^{1/z}\)). However, this
interpretation has been criticized in this and other nonequilibrium
models \cite{hgag93,BeHo-1,BeHo-2}. On the other hand, the second symmetry
essentially tells us that in 1D, the nonlinear (KPZ) term is not
operative at long times.

Even when recognizing the interesting analytical properties of the KPZ
equation, it is clear that investigating the behavior of its
solutions requires the (stochastic) numerical integration of a
discrete version. Such an approach has been used ,e.g., to obtain the
critical exponents in one and more spatial dimensions
\cite{BecCur,MoserWolf-Discr3d,Scalerandi-etal,NewmanBray-Discr,Appert,FoTo,LamShin}.
Although a
pseudo-spectral spatial discretization scheme has been recently
introduced \cite{GiadaGiacomettiRossi,gacl07}, real-space discrete versions
of Eq. (\ref{eq-000}) are still used for numerical simulations
\cite{taea04,majy07}. One reason is their relative ease of
implementation and of interpretation in the case of non-homogeneous
substrates, for example  a quenched impurity distribution
\cite{dlea09}.

\subsection*{Consistency}

Here, we use the standard, nearest-neighbor discretization prescription
as a benchmark to elucidate the constraints to be obeyed by any
spatial discretization scheme, arising from the mapping between the
KPZ and the diffusion equation (with multiplicative noise) through
the Hopf--Cole transformation.

The standard spatially discrete version of Eq. (\ref{eq-0002}) is
\begin{equation}\label{eq:3.01}
\dot{\phi}_j=
\frac{\nu}{a^2}\,\left(\phi_{j+1}-2\phi_j+\phi_{j-1}\right)+
\frac{\lambda \sqrt{\gamma} }{2\nu}\phi_j\xi_j,
\end{equation}
with \(1\le j\le N\equiv0\), because of the assumed periodic b.c. (the
implicit sum convention is not meant in any of the discrete
expressions). Here \(a\) is the lattice spacing. Then, using the
discrete version of Hopf--Cole transformation
$\phi_j(t)=\exp\left[\frac{\lambda}{2\nu}h_j(t)\right],$ we get
\begin{equation}\label{eq:3.03}
\dot{h}_j=\frac{2\nu^2}{\lambda a^2}\left(\mathrm{e}^{\delta_j^+a}
+\mathrm{e}^{\delta_j^-a}-2\right)+ \sqrt{\gamma}\,\xi_j,
\end{equation}
with \(\delta_j^\pm\equiv\frac{\lambda}{2\nu a}(h_{j\pm1}-h_j)\). By
expanding the exponentials up to terms of order \(a^2\), and
collecting equal powers of \(a\) (observe that the zero-order
contribution vanishes) we retrieve
\begin{eqnarray}
\dot{h}_j &=& \frac{\nu}{a^2}\left(h_{j+1}-2h_j+h_{j-1}\right)\nonumber \\
& & + \frac{\lambda}{4\,a^2}\left[(h_{j+1}-h_j)^2+(h_j-h_{j-1})^2\right]\nonumber\\
& & + \sqrt{\gamma}\,\xi_j. \label{eq:3.04}
\end{eqnarray}
As we can see, the first and second terms on the r.h.s. of Eq.\
(\ref{eq:3.04}) are \emph{strictly} related by virtue of the
Hopf-Cole transformation. In other words, the discrete form of the
Laplacian in Eq. (\ref{eq:3.01}) constrains the discrete form of the
nonlinear term in the transformed equation. Later we return, in
another way, to the tight relation between the discretization of both
terms. Known proposals \cite{LamShin} fail to comply with this
natural requirement.

An important feature of the Hopf--Cole transformation is that it is
\emph{local}, i.e., it involves neither spatial nor temporal
transformations. An effect of this feature is that the discrete form
of the Laplacian is the same, regardless of whether it is applied to
\(\phi\) or \(h\).

The aforementioned criterion dictates the following discrete form
for \(\mathcal{F}[\phi]\) (the one just before Eq. (\ref{eq-001})), thus a
Lyapunov function for any finite \(N\)
\begin{eqnarray}\label{eq:3.05}
\mathcal{F}[\phi]&=&\frac{\nu}{2}\sum_{j=1}^Na\left((\partial_x\phi)^2
\right)_j\nonumber\\
&=&\frac{\nu}{4a}\sum_{j=1}^N
\left[(\phi_{j+1}-\phi_j)^2+(\phi_j-\phi_{j-1})^2\right].\nonumber \\
& &
\end{eqnarray}
It is a trivial task to verify that the Laplacian is
\((\partial_x^2\phi)_j=-a^{-1}\partial_{\phi_j}\mathcal{F}[\phi]\).
Now, the obvious fact that this functional can also be written as
\(\mathcal{F}[\phi]=\frac{\nu}{2\,a}\sum_{j=1}^N(\phi_{j+1}-\phi_j)
^2\) illustrates a fact that for a more elaborate
discretization requires explicit calculations: the Laplacian does
not \emph{uniquely} determine the Lyapunov function \cite{Nos-1,Nos-2,Nos-3}.

Equation (\ref{eq:3.03}) has also been
written in \cite{NewmanBray-Discr}, although with different goals than ours.
Their interest was to analyze the strong coupling limit via mapping
to the directed polymer problem.

\subsection*{An accurate consistent discretization}

Since the proposals of \cite{LamShin} already involve
next-to-nearest neighbors, one may seek for a prescription that
minimizes the numerical error. An interesting choice for the
Laplacian is \cite{abst65}
\begin{equation}\label{eq:3.06}
\frac{1}{12\,a^2}\left[16(\phi_{j+1}+\phi_{j-1})
-(\phi_{j+2}+\phi_{j-2})-30\,\phi_j\right],
\end{equation}
which has the associated discrete form for the KPZ term
\begin{eqnarray}
(\partial_x\phi)^2&=&\frac{1}{24\,a^2} \left\{16\left[(\phi_{j+1}
-\phi_j)^2\right.\right.\nonumber \\
& & \,\,\,\,\,\,\,\,\,\,\,\,\,\, \left. +(\phi_j-\phi_{j-1})^2\right]\nonumber\\
&&-\left.\left[(\phi_{j+2}-\phi_j)^2+(\phi_j-\phi_{j-2})^2\right]
\right\}\nonumber \\
& & \,\,\,\,\,\,\,\,\,\,\, +\mathcal{O}(a^4). \label{eq:3.07}
\end{eqnarray}
Replacing this into the first line of Eq. (\ref{eq:3.05}), we
obtain Eq. (\ref{eq:3.06}). Since this discretization scheme
fulfills the consistency conditions, it is accurate up to
\(\mathcal{O}(a^4)\) corrections, and its prescription is not more
complex than other known proposals, we expect that it will be the
convenient one to use when high accuracy is required in numerical
schemes \cite{Nos-1,Nos-2,Nos-3}.

\subsection*{Relation with the Lyapunov functional}

In Sect. III we have indicated the form of the NEP for KPZ, and the
way in which the functionals \(\mathcal{F}[\phi]\) and
\(\mathcal{F}[h]\) are related \cite{wio-01}. According to the
previous results, we can write the discrete version of Eq.\
(\ref{eq-001}) as
\[\mathcal{F}[h]=\frac{\lambda^2}{8\nu}\frac{1}{2\,a}\sum_j
e^{\frac{\lambda}{\nu}h_j}\left[(h_{j+1}-h_j)^2 \right.\]
\[\,\,\,\,\,\,\,\,\,\, \left. +(h_j-h_{j-1})^2\right].\]
Introducing this expression into
$\partial _t h_j = \Gamma_j \frac{\delta \mathcal{F}[h]}{\delta h_j}$,
and through a simple algebra, we obtain Eq. (\ref{eq:3.04}). This
reinforces our previous result, and moreover indicates that the
discrete variational formulation naturally leads to a consistent
discretization of the KPZ equation.

\subsection*{The fluctuation--dissipation relation}

This relation is, together with Galilean invariance, a
fundamental symmetry of the one-dimensional KPZ equation. It is
clear that both symmetries are recovered when the continuum limit is
taken in any reasonable discretization scheme. Thus, an accurate
enough partition must yield suitable results.

The stationary probability distribution for the KPZ problem in 1D is
known to be \cite{HHZ,BarSta}
\[\mathcal{P}_\mathrm{stat}[h]\sim\exp\left\{-\frac{\nu}{2\,
\gamma}\int d x\left(\partial_xh\right)^2\right\}.\] For the
discretization scheme in Eq. (\ref{eq:3.04}), this is
\begin{equation}\label{eq:3.08}
\sim\exp\left\{\frac{\nu}{2\varepsilon}\frac{1}{2a}
\sum_j\left[(h_{j+1}-h_j)^2+(h_j-h_{j-1})^2\right]\right\}.
\end{equation}
Inserting this expression into the stationary Fokker--Planck
equation, the only surviving term has the form
\begin{eqnarray}\label{eq:3.09}
\frac{1}{2a^3}\sum_j& &\left[(h_{j+1}-h_j)^2\right.+\left.(h_j-h_{j-1})^2\right]\nonumber\\
& &\times\left[h_{j+1}-2h_j+h_{j-1}\right].
\end{eqnarray}
The continuum limit of this term is \(\int d
x\left(\partial_xh\right)^2\partial_x^2h\), that is identically zero
\cite{HHZ,BarSta}. A numerical analysis of Eq. (\ref{eq:3.09})
indicates that it is several orders of magnitude smaller than the
value of the exponents' pdf [in Eq. (\ref{eq:3.08})], and typically
behaves as \(\mathcal{O}(1/N)\), where \(N\) is the number of
spatial points used in the discretization. Moreover, it shows an
even faster approach to zero if expressions with higher accuracy
[like Eqs. (\ref{eq:3.06}) and (\ref{eq:3.07})] are used for the
differential operators. In addition, when the discrete form of
\((\partial_xh)^2\) from \cite{LamShin} is used together with its
consistent form for the Laplacian, the fluctuation--dissipation
relation \textbf{is not} exactly fulfilled. This indicates that the
problem with the fluctuation--dissipation theorem in \(1+1\),
discussed in \cite{GiadaGiacomettiRossi,LamShin} can be just
circumvented by using more accurate expressions.

\subsection*{Galilean invariance}

This invariance means that the transformation
\begin{equation}\label{eq:3.10}
x\to x-\lambda vt,\quad h\to h+vx,\quad F\to F-
\frac{\lambda}{2}v^2,
\end{equation}
where \(v\) is an arbitrary constant vector field, leaves the KPZ
equation invariant. The equation obtained using the classical
discretization
\begin{equation}\label{eq:3.11}
\partial_xh\to\frac{1}{2\,a}(h_{j+1}-h_{j-1}),
\end{equation}
is invariant under the discrete Galilean transformation
\begin{equation} \label{eq:3.12}
ja\to ja-\lambda vt,\quad h_j\to h_j+vja,\quad F\to F-
\frac{\lambda}{2}v^2.
\end{equation}
However, the associated equation is known to be numerically unstable
\cite{NewmanBray-Discr}, at least when \(a\) is not small enough. Besides,
Eq. (\ref{eq:3.04}) is not invariant under the discrete
Galilean transformation. In fact, the transformation \(h\to h+vja\)
yields an excess term which is compatible with the gradient
discretization in Eq. (\ref{eq:3.11}); however, this discretization
does not allow to recover the quadratic term in Eq.\
(\ref{eq:3.04}), indicating that this finite-difference scheme is
not Galilean-invariant.

Since Eq. (\ref{eq:3.01}) is invariant under the transformation
indicated in Eq. (\ref{eq:3.12}), it is the nonlinear Hopf--Cole
transformation (within the present discrete context) which is
responsible for the loss of Galilean invariance. Note that these
results are independent of whether we consider this discretization
scheme or a more accurate one.

Galilean invariance has always been associated with the exactness of
the one-dimensional KPZ exponents, and with a relation that connects
the critical exponents in higher dimensions \cite{medina}. If the
numerical solution obtained from a finite-difference scheme as
Eq. (\ref{eq:3.04}), which is not
Galilean invariant, \emph{yields the well known critical exponents},
this will be an indicative that Galilean invariance is not strictly
necessary to get the KPZ universality class. The numerical results
presented in \cite{Nos-1,Nos-2,Nos-3} clearly show that this is the case.

%---------------------------------------
We will not discuss here the simulation procedure but only
indicate that to make the simulations we introduced a discrete
representation of \(h(x,t)\) along the substrate direction \(x\) with lattice
spacing \(a = 1\), and that a standard second-order Runge--Kutta
algorithm (with periodic boundary conditions) was employed
(see  \cite{Toral}).
In \cite{Nos-1,Nos-2,Nos-3} it was shown that all the cases
(consistent or not) exhibit the same critical exponents. Moreover, we
want to note that the discretization used in Refs. \cite{LamShin},
which also violates Galilean invariance, yields the same critical
exponents too. Additionally, stochastic differential equations which
are not explicitly Galilean invariant have been shown to obey the
relation $\alpha + z = 2$ (\cite{NiCuCa}, see also next section).
Hence, our numerical analysis indicates that there are discrete
schemes of the KPZ equation which, even not obeying Galilean
invariance, show KPZ scaling.
%---------------------------------------

The moral from the present analysis is clear: due to the locality of
the Hopf--Cole transformation, the discrete forms of the Laplacian
and the nonlinear (KPZ) term cannot be chosen independently;
moreover, the prescriptions should be the same, regardless of the
fields they are applied to. For further details we refer to
 \cite{Nos-1,Nos-2,Nos-3}.

\section{Renormalization-group analysis for Fluctuations}

In section III we have built a gradient flow counterpart of the deterministic
KPZ equation. In this section we consider the corresponding stochastically
forced gradient flow
\begin{equation}
\partial_t u = -\frac{\delta \mathcal{I}}{\delta u}+ \xi(x,t),
\end{equation}
with the density indicated in Eq.~(\ref{nep-17}).
We obtain the KPZW equation, which is the following SPDE
\begin{equation}\label{kpzp}
\partial_t u = \nu \nabla^2 u - \frac{\lambda}{6} (\nabla u)^2
-\frac{\lambda}{3} u \nabla^2 u + \xi(x,t).
\end{equation}
Our present goal will be to analyze the scaling behavior of the fluctuations of the
solution to this equation.

Since Eq.~(\ref{kpzp}) is nonlinear, we focus on a perturbative technique. We
choose the dynamic renormalization group as employed in~\cite{fns,medina}. Employing
this method, we find at one-loop order the following flow equations \cite{Nos-PLA}
\begin{eqnarray}
\label{flow1}
\frac{d \lambda}{d \ell} &=& \lambda (\alpha + z -2), \\
\label{flow2}
\frac{d \nu}{d \ell} &=& \nu \left(z-2 - \frac{1}{36} \frac{\lambda ^2 D}{\nu^3} K_d \frac{1-d}{d} \right), \\
\label{flow3} \frac{d \gamma}{d \ell} &=& \gamma \left( z-d-2 \alpha
+\frac{K_d}{72} \frac{\lambda^2 \gamma}{\nu^3} \right),
\end{eqnarray}
where $K_d = S_d/(2 \pi)^d$, $S_d=2 \pi^{d/2}/\Gamma (n/2)$
is the surface area of the $d-$dimensional unit sphere, and $\Gamma$ is the gamma function.
We find that the coupling constant $\bar{g} :=K_d\lambda^2 \gamma/\nu^3$ obeys the
one-loop differential
equation
\begin{equation}\label{coupconst}
\frac{d \bar{g}}{d \ell}= (2-d) \bar{g} + \frac{6-5d}{72d}
\bar{g}^2,
\end{equation}
revealing that the critical dimension of this model is $d_c=2$ as could be anticipated by
means of power counting. For $d > 2$ the coupling
constant approaches zero exponentially fast in the scale $\ell$; for $d=2$, this approach
is algebraic. So for these dimensions one expects the large-scale space-time properties of
Eq.~(\ref{kpzp}) to be dominated by its linear counterpart (up to marginal corrections in
$d=2$). In $d=1$, the coupling constant runs to infinity for finite $\ell$, suggesting the
presence of a non-perturbative fixed point (as the one in the KPZ equation for $d=2$).

The values of the critical exponents which yield scale invariance can be formally calculated
by identifying with zero the right hand sides of Eqs.~(\ref{flow1})--(\ref{flow3}). We get
\begin{eqnarray}
\alpha&=&\frac{2(2-d)(1-d)}{6-5d},\\
z&=&\frac{12-10d-2(2-d)(1-d)}{6-5d},
\end{eqnarray}
which in particular obey the relation $\alpha+z=2$ in any dimensionality, despite the fact
that Eq.~(\ref{kpzp}) does not obey any sort of Galilean invariance. We note that in both
$d=1$ and $d=2$ it is $\alpha = 0$ and $z=2$, whereas $\alpha$ becomes negative in higher
dimensions. Hence, in all dimensions, the exponent $\alpha$ indicates that the interface
is either flat or at most marginally rough. The values for $d=1$ make both diffusion and
nonlinearity in Eq.~(\ref{kpzp}) invariant under the scale transformation
$\{x,t,u\}\to\{bx,b^zt,b^\alpha u\}$, as far as $b>1$. In this case, the noise grows with
the scale (a fact that might explain the growth of the coupling constant in the renormalization
group flow). In $d=2$, the exponents are those of the linear equation. An interesting result is
that for $d=0$, the exponents become those of the KPZ equation: $\alpha=2/3$ and $z=4/3$,
although this limit is highly singular for Eq.~(\ref{coupconst}). Of course, these results have
been obtained by means of a perturbative dynamic renormalization group and could be modified by
non-perturbative contributions.
One possible path to study such a possibility could be to adapt some \emph{non perturbative
renormalization group} techniques used for KPZ \cite{canet} to the present KPZW case.

Among all the results in this section, we would like to highlight the one given by Eq.~(\ref{flow1}).
We recall that the RG analysis of the KPZ equation yields non-renormalization of the vertex and
renormalization of propagator and noise. Our variational equation yields exactly the same result.
Vertex non-renormalization at one-loop order is expressed by Eq.~(\ref{flow1}). The origin of this
result is analogous to that of its equivalent in the KPZ equation: three non-vanishing Feynman
diagrams contribute to vertex renormalization, but they cancel out each other~\cite{kpz} (a fact
that has been traditionally attributed to the Galilean invariance of the KPZ equation). Here we
have shown that the same result appears in a SPDE that is not even invariant under the
translation $u \to u+$constant.

\section{Stability}

We have carried out the NEP expansion about a constant solution of the KPZ equation and
found that constants are still solutions to KPZW (Eq.~(\ref{nep-18})). In this section we will
study the linear stability of such solutions. We start considering the solution
\begin{equation}
u(x,t) = c + \epsilon \upsilon(x,t),
\end{equation}
where $c$ is an arbitrary constant and $\epsilon$ is the small parameter.
Substituting in Eq.~(\ref{nep-18}), we find
\begin{equation}
\partial_t \upsilon = \frac{3 \nu -  \lambda c }{3} \, \nabla^2 \upsilon
,
\end{equation}
at first order in $\epsilon$. So $\upsilon$ obeys a diffusion equation whose diffusion
constant depends on $c$. For $c<3\nu/2\lambda$, the diffusion constant is positive and
correspondingly the constant solution is linearly stable. For $c>3\nu/2\lambda$, the
diffusion constant is negative and consequently the constant solution is unstable. Furthermore, in this case the problem becomes linearly ill posed.

Since for large values of $c$, the problem becomes linearly ill posed, numerical
solutions are not available. In order to solve this disadvantage, we could
include a higher order term in our problem. We concentrate on the gradient flow
\begin{equation}
\partial_t u = -\frac{\delta \mathcal{J}}{\delta u}+ \xi(x,t),
\end{equation}
with density
\begin{eqnarray}
\mathcal{J}[u] & = & \frac{\nu}{2} \int \,\mathrm{d} x \, (\nabla u)^2 -
\frac{\lambda}{6} \int \,\mathrm{d} x \,\, u (\nabla u)^2 \nonumber\\
& & \,\,\,\,\,\,\,\,\,\,\,\,\,\, + \frac{\mu}{2} \int\,\mathrm{d} x \, (\nabla^2 u)^2,
\end{eqnarray}
leading to the following equation
\begin{equation}
\partial_t u = \nu \nabla^2 u - \frac{\lambda}{6} (\nabla u)^2
-\frac{\lambda}{3} u \nabla^2 u - \mu \nabla^4 u + \xi(x,t).
\end{equation}
Note that the deterministic counterpart of this fourth-order equation
can be considered as a variational version of the Kuramoto-Sivashinsky equation.
It is worth remarking that this ad hoc construction resembles the one that, as
indicated in~\cite{wio-01}, could more formally be obtained by considering the
expansion of a nonlocal, short range interaction.

The regime of linear stability/instability of this equation is identical to that
of Eq.~(\ref{kpzp}) but in this case, the problem is always linearly well posed.
Furthermore, the term proportional to $\mu$ is presumably irrelevant in the large
spatiotemporal scale (as simple power counting of the linear terms reveals) so
the results of the previous RG analysis could possibly hold for this case too.
Anyway, due to the presence of the deterministic instability, further analysis are
needed in order to assure this (note that both linear terms in the equation are
stabilizing and that this instability has its origin in the vertex structure).

\section{Crossover: a path integral point of view}

Another recently discussed related aspect \cite{Nos-5} is based in a
path-integral Monte Carlo-like method for the numerical evaluation of the mean
rugosity and other typical averages whose approach, which radically differs from
one introduced before \cite{FoRe}, exploits some of our previous results
\cite{Nos-1,Nos-2,Nos-3}. Here we limit ourselves to quote the temporally
($\mu$) \emph{and} spatially ($j$) discrete form of the ``stochastic action''
\begin{eqnarray}\label{eq:2}
\mathbf{S}[h]&=&\frac1{2\tau}\sum_{j,\mu}\left\{h_{j,\mu +1}-h_{j,\mu} \right. \nonumber\\
& & \left.-\tau[\alpha\mathbf{L}_{j,\mu+1}+(1-\alpha)\mathbf{L}_{j,\mu}]\right\}^2\nonumber\\
& &-2\nu\alpha Nt \nonumber \\
& & -\tau\alpha\frac{\lambda}2\sum_{j,\mu}[h_{j+1,\mu}-2h_{j,\mu}+h_{j-1,\mu}],\nonumber\\
\end{eqnarray}
and briefly discuss the obtained numerical results. $\tau$ is the time step,
$0<\alpha<1$ a time-discretization parameter meant to be fixed for explicit
calculation \cite{tirap,wio-ip}, and $\mathbf{L}_{j,\mu}$ the ``stochastic
Lagrangian''
\begin{eqnarray}\label{eq:3}
\mathbf{L}_{j,\mu}&=&\nu \left(h_{j+1,\mu} - 2 h_{j,\mu} + h_{j-1,\mu} \right)\nonumber\\
& & + \frac{\lambda}{4} \left[(h_{j+1,\mu}-h_{j,\mu})^2 \right.\nonumber\\
& & \left. + (h_{j,\mu}-h_{j-1,\mu})^2\right].
\end{eqnarray}

\begin{figure}%[th]
\begin{center}
\includegraphics[width=0.48\textwidth]{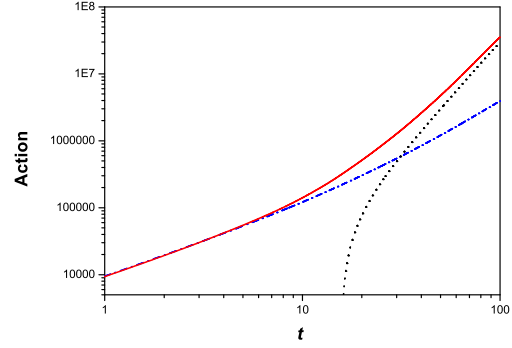}
\end{center}
\caption{Crossover-like behavior from EW to KPZ regime for $\lambda=1$,
on a lattice of
1028 sites ($\nu=D=1$). Red solid line: KPZ action; blue dash-dotted
line: EW action; black-dotted line: difference.}\label{fig:1}
\end{figure}

\begin{figure}%[th]
\begin{center}
\includegraphics[width=0.48\textwidth]{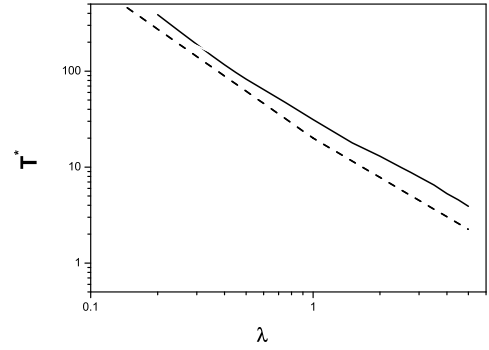}
\end{center}
\caption{Same data as the previous figure. Solid line: time
$T^*$ \emph{vs.} $\lambda$; dashed line: trend for $T^*\sim\lambda^{-1.35}$,
included for comparison.}\label{fig:2}
\end{figure}

Figure \ref{fig:1} shows the crossover-like behavior from the Edwards--Wilkinson (EW)
regime to the KPZ one. We take as estimator of such a transition the time at which the
difference (dotted black line) between KPZ (red solid curve) and EW actions (blue
dash-dotted line) crosses the EW one (it grossly coincides with the time at which
the asymptotes cross). This estimator numerically agrees neither with the results
in \cite{FoTo} (where a value of $\phi \sim 4$
was found) nor with the one in \cite{GGG} (with $\phi\sim 3$, but corresponding to a 2D
case). In Fig. \ref{fig:2} we have plotted the dependence of this estimator on
$\lambda$. For comparison, we have also included the trend for $\lambda^{-\phi}$
with $\phi=1.35$ (dotted line). Preliminary results for $\lambda > 7$ seem to
indicate a marked change in the value of $\phi$, maybe a hint that the
system is entering a \emph{strong coupling} region \cite{FoRe}.

\subsection*{Short-time propagator}
Our aim here is to work out a variant of the method introduced in \cite{Nos-5}
by exploiting the first form of Lyapunov functional found in \cite{wio-01},
namely Eq. (\ref{nep-01}), that leads us to Eq. (\ref{kpz-var}), the full KPZ equation.

Whereas Eq. (\ref{eq:2}) is valid whatever the value of $\tau$, we now seek for a
simpler expression valid for $\tau \ll 1$. This idea parallels in some sense other
studies in the literature \cite{FoRe}, but here we exploit the functional
$\mathcal{F}[h]$ of Eq. (\ref{eq-001}). We denote as
$\{h\}=(h_{1,\mu},h_{2,\mu},\ldots,h_{j,\mu},\ldots,h_{N,\mu})$ the interface
configuration at time $\mu$. The transition pdf between patterns $h_0$ at
$t_0$ and $h_f$ at $t_f$ can be written as
\begin{eqnarray}\label{eq:6}
&&P(\{h_f\},t_f|\{h_0\},t_0)=\nonumber\\
&&\,\,\,\,\,\,\, \int\mathcal{D}[h]\exp\left(-\frac{1}\gamma\int_{t_0}^{t_f}\mathcal{L}[h,\dot{h}]\right),
\end{eqnarray}
with
\begin{eqnarray}\label{eq:7}
\mathcal{L}[h,\dot{h}]&=&\frac{1}2\int_0^Ldx
\left[\left(\partial_th+\Gamma[h]\frac{\delta}{\delta h}\mathcal{F}[h]\right)^2 \right.\nonumber\\
& & \,\,\,\,\, \left.+ \alpha \frac{\delta}{\delta h}\left(\Gamma [h]\frac{\delta}{\delta h}\mathcal{F}[h]\right)\right],
\end{eqnarray}
whose discrete form is given by Eq. (\ref{eq:3}). A key observation is the
(temporally and spatially) ``diagonal'' character of Eq. (\ref{eq-002}), highlighted
in its discrete version
\begin{equation}\label{eq:6-2}
\dot{h}_j(t)=-\Gamma_j\frac{\delta\mathcal{F}}{\delta h_j}+\sqrt{\gamma}\,\xi_j(t).
\end{equation}
Guided by Eq. (\ref{eq:3}), we propose the following form of $P(\{h_f\},t_f|\{h_0\},t_0)$
for $\tau \ll 1$, or short-time propagator (STP)
\begin{eqnarray}\label{eq:19}
&& P(h_f,\tau|h_0,0) = \nonumber\\
&& \int_{h_0}^{h_f}\mathcal{D}[h]
e^{\left[-\frac1{2\gamma}\int_0^\tau ds
\int_0^Ldx\left(\partial_th+\Gamma\frac{\delta\mathcal{F}}{\delta h}\right)^2\right]}\nonumber\\
&&\approx \exp\Bigl\{-\frac{\tau}{2 \gamma}\int_0^L\,dx \Bigr.\nonumber\\
&& \,\,\,\,\,\,\,\,\,\,\,\,\,\,\, \left[\left(\frac{h_f-h_0}{\tau}
+\frac{1}{2}\left[\Gamma_f\frac{\delta\mathcal{F}}{\delta h_f}
+\Gamma_0\frac{\delta\mathcal{F}}{\delta h_0}\right]\right)^2 \right]\Bigr\}.\nonumber\\
%&&\,\,\,\,\,\,\,\,\,\Bigl.\left.-\alpha\frac{\delta}{\delta h}\left(\Gamma[h]\frac{\delta}{\delta h}\mathcal{F}[h]\right)\right]\Bigr\}.
&&
\end{eqnarray}
Here, for simplicity, we have chosen a discretization with $\alpha = 0.$
As it is well known \cite{tirap,wio-ip}, the Jacobian of the transformation
from the noise variable to the height variable depends on $\alpha.$ With
this choice, the Jacobian results equal to 1.

Incidentally, the form in Eq. (\ref{eq:19}) coincides with the discretization used
in \cite{FoRe} for determining the least-action trajectory. The ``quasi-Gaussian"
character of this STP is better evidenced in the following approximate form
\begin{eqnarray}\label{eq:20}
&&P(h_f,t_f=\tau|h_0,t_0=0) \sim e^{\left[-\frac{1}{2\gamma\tau}\int_0^Ldx
\left(h_f-h_0\right)^2\right]}\nonumber\\
&&\,\,\,\,\,\,\times\Bigl\{1-\frac{1}{2\gamma}\int_0^Ldx
\left[(h_f-h_0)\frac{1}{2}\left(\Gamma_f\frac{\delta\mathcal{F}}{\delta h_f} \right.\right.\Bigr.\nonumber\\
&&\,\,\,\,\,\,\,\,\,\,\,\,\Bigl. \left. \left. +\Gamma_0\frac{\delta\mathcal{F}}{\delta h_0}\right)
+  \mathcal{O}(\tau)\right]\Bigr\}
\end{eqnarray}
where the exponential term has been separated out since it is of order $\tau^{-1}$,
whereas the following two are of order $\tau^0$ and $\tau^1$, respectively (of lesser
weight and negligible respectively, in the limit $\tau\to0$). It is worth remarking
that the term that could come from the Jacobian is also of order $\tau^1$.

It is easy to check that we can recover the known FPE from the proposed form of STP
(adopting $\alpha=0$ for simplicity). We will not reiterate this calculation here.
An immediate result of this
form is that at very short times, behavior of the Edwards--Wilkinson type is obtained
$$\sqrt{\langle h^2\rangle}\approx\tau^{\frac{1}2}.$$

\section{Conclusions}

Herein, in addition to reviewing some recent results \cite{Nos-1,Nos-2,Nos-3,Nos-PLA,Nos-5},
we have furthered the study in~\cite{wio-01}, where it was shown that the
deterministic KPZ equation admits a Lyapunov functional, and a (formal) definition
of a \emph{nonequilibrium potential} was introduced. We have carried out a
Taylor expansion of such a nonequilibrium potential, what led us to a different
equation of motion than the KPZ one, the KPZW which is an exact gradient flow and has an
explicit density. In particular, it has a lower degree of symmetry: it is neither
Galilean invariant, nor even translational invariant. The critical exponents
determining its scaling properties were obtained through a one-loop
dynamic renormalization group analysis. These exponents fulfill the same scaling
relation as the KPZ equation, $\alpha+z=2$, traditionally attributed to the Galilean
invariance of the latter. The fact that the same scaling relation arises in a SPDE
(i.e., the KPZW) that is not only non-Galilean invariant but even non-invariant under
the translation $u \to u+$constant supports recent theoretical and numerical results
indicating that Galilean invariance does not necessarily play the relevant role
previously assumed in defining the universality class of the KPZ equation and
different nonequilibrium models~\cite{McComb,BeHo-1,BeHo-2,NiCuCa,Nos-1,Nos-2,Nos-3}.

We have, moreover, analyzed the stability properties of the solutions to the present
equation, finding the threshold condition for the appearance of diffusive instabilities,
which indicates that in this case the problem becomes linearly ill posed. After
considering the simplest way to correct such an ill-posed problem, we have met a kind of
Kuramoto--Sivashinsky equation, resembling the one that, as indicated in~\cite{wio-01},
could be obtained by considering a nonlocal, short range interaction. This equation
has an exact gradient flow structure with an explicit density. Furthermore, when
subject to stochastic forcing, its scaling properties could be formally described
by the same critical exponents because the stabilizing term is irrelevant in the
large scale from a dimensional analysis viewpoint.

Exploiting some elements of a path integral description of the problem, we have
also shown what seems to be a simple form of viewing and studying the crossover from
the EW to the KPZ regimes.

The present review-like study aims to open new points of view on, as well as
alternative routes to study, the KPZ problem.
Among the many aspects to be further studied, an interesting one is
to test the (kind) of stability of the recently found exact solutions
\cite{SaSp,SaSp2,AmCoQu,CaLeD} by exploiting the indicated form of the NEP.

\begin{acknowledgements}
Financial support from MINECO (Spain) is especially acknowledged, through
Projects PRI-AIBAR-2011-1323 (which enabled international cooperation),
FIS2010-18023 (HSW and CE) and RYC-2011-09025 (CE). Also acknowledged is
the support from CONICET, UNC (JAR) and UNMdP (RRD) of Argentina.
Collaboration with M.S. de la Lama and E. Korutcheva during different stages
of this research is highly appreciated.
\end{acknowledgements}

\end{document}